\documentclass[final,3p,twocolumn,times,9pt]{elsarticle}

\usepackage{amssymb}
\usepackage{amsmath}
\usepackage{xcolor}
\usepackage{xspace}
\usepackage{cuted}
\usepackage{booktabs}
\usepackage{comment}
\usepackage{hyperref}
\usepackage{cleveref}
\usepackage[squaren]{SIunits}

\usepackage{amsmath}
\usepackage{amssymb}
\usepackage{commath}
\usepackage{array}
\usepackage{calc}
\usepackage{longtable}
\usepackage{multirow}
\usepackage{physics}
\usepackage{tensor}
\usepackage{upgreek}

\usepackage{graphicx}
\graphicspath{{figures/}}
\usepackage{xspace}

\newcommand{\Comix}{C\protect\scalebox{0.8}{OMIX}\xspace}

\newcommand{\Geneva}{G\protect\scalebox{0.8}{ENEVA}\xspace}

\newcommand{\MCFM}{M\protect\scalebox{0.8}{CFM}\xspace}

\newcommand{\Pythia}{P\protect\scalebox{0.8}{YTHIA}\xspace}
\newcommand{\Rambo}{R\protect\scalebox{0.8}{AMBO}\xspace}
\newcommand{\Sherpa}{S\protect\scalebox{0.8}{HERPA}\xspace}
\newcommand{\Vincia}{V\protect\scalebox{0.8}{INCIA}\xspace}

\newcommand{\MCatNLO}{M\protect\scalebox{0.8}{C}@N\protect\scalebox{0.8}{LO}\xspace}
\newcommand{\Powheg}{P\protect\scalebox{0.8}{OWHEG}\xspace}
\newcommand{\minlo}{MiNLO\xspace}
\newcommand{\minnlops}{MiNNLO$_{\mathrm{PS}}$\xspace}
\newcommand{\unlops}{UNLOPS\xspace}
\newcommand{\untwolops}{UN$^2$LOPS\xspace}

\newcommand{\D}{\ensuremath{\,\text{d}}}

\newcommand{\Order}[1]{\ensuremath{\mathcal{O}\left(#1\right)}}

\newcommand{\alphaS}{\ensuremath{\alpha_{\mathrm{s}}}}

\newcommand{\tc}{\ensuremath{t_\mathrm{c}}}
\newcommand{\tzero}{\ensuremath{t_0}}

\newcommand{\mursq}{\ensuremath{\mu}_\mathrm{R}^2}

\newcommand{\CalC}{{\mathcal{C}}}

\newcommand{\CalS}{{\mathcal{S}}}

\newcommand{\CalM}{{\mathcal{M}}}
\newcommand{\UA}{{\mathrm{A}}}
\newcommand{\UB}{{\mathrm{B}}}
\newcommand{\UV}{{\mathrm{V}}}
\newcommand{\UR}{{\mathrm{R}}}
\newcommand{\UVV}{{\mathrm{V\kern-0.15em V}}}
\newcommand{\URV}{{\mathrm{R\kern-0.15emV}}}
\newcommand{\URR}{{\mathrm{R\kern-0.15em R}}}
\newcommand{\US}{{\mathrm{S}}}
\newcommand{\UT}{{\mathrm{T}}}

\newcommand{\UI}{{\mathrm{I}}}
\newcommand{\avg}[1]{\ensuremath{\langle #1 \rangle}}
\renewcommand{\abs}[1]{\ensuremath{\left\vert #1 \right\vert}}

\newcommand{\nlo}{\mathrm{NLO}}

\biboptions{numbers,sort&compress}

\journal{Physics Letters B}

\begin{document}

\begin{frontmatter}

\title{Towards NNLO+PS Matching with Sector Showers}

\author[label1]{John M Campbell}
\ead{johnmc@fnal.gov}
\author[label1]{Stefan H\"oche}
\ead{shoeche@fnal.gov}
\author[label4,label2]{Hai Tao Li}
\ead{haitao.li@northwestern.edu}
\author[label3]{Christian T Preuss}
\ead{christian.preuss@monash.edu}
\author[label3]{Peter Skands}
\ead{peter.skands@monash.edu}
\address[label1]{Fermi National Accelerator Laboratory, Batavia, IL, 60510, USA}
\address[label4]{HEP Division, Argonne National Laboratory, Argonne, Illinois 60439, USA}
\address[label2]{Department of Physics \& Astronomy, Northwestern University, Evanston, IL 60208, USA}
\address[label3]{School of Physics and Astronomy, Monash University, Wellington Road, Clayton, VIC-3800, Australia}

\begin{abstract}
We outline a new technique for the fully-differential matching of final-state parton showers to NNLO calculations, focussing here on the simplest case of leptonic collisions with two final-state jets. The strategy is facilitated by working in the antenna formalism, making use of NNLO antenna subtraction on the fixed-order side and the sector-antenna framework on the shower side. As long as the combined real-virtual and double-real corrections do not overcompensate the real-emission term in the three-jet region, negative weights can be eliminated from the matching scheme. We describe the implementation of all necessary components in the \Vincia antenna shower in \Pythia~8.3.
\end{abstract}


\end{frontmatter}

\section{Introduction}
\label{sec:int}
To date it is possible to perform most collider-physics studies with  fully-differential NLO+PS matching thanks to two general, well-developed, and widely applied procedures: \MCatNLO \cite{Frixione:2002ik} and \Powheg \cite{Nason:2004rx}. 
By fully differential matching, we understand that the matching is done point by point in both the Born- and real-emission phase spaces, with a parton shower that reflects the correct singular structure of the fixed-order calculation. In this sense, fully-differential matching requires the fixed-order expansion of the shower to develop the same singularities as the fixed-order calculation up to the matched order. At NLO, this is achieved by parton showers that exponentiate terms reducing to the universal DGLAP kernels in any collinear limit and the eikonal factor in soft limits, with the colour dependence in the soft limit requiring special attention~\cite{Hoeche:2011fd}. As of today, no fully-differential matching method obeying these criteria is available at NNLO, although significant progress on including higher-order corrections to parton showers has been made \cite{Li:2016yez,Hoche:2017iem,Hoche:2017hno,Dulat:2018vuy}.

Existing NNLO+PS matching methods either extend existing merging schemes or utilise analytical resummation for the transition between the fixed-order and parton-shower realms.
Examples of the first kind are \untwolops \cite{Hoche:2014uhw}, which extends the \unlops \cite{Lonnblad:2012ix} scheme to the second order, and \minnlops \cite{Monni:2019whf,Re:2021vcb} as well as other extensions~\cite{Hamilton:2013fea,Karlberg:2014qua,Astill:2016hpa,Astill:2018ivh,Hamilton:2012rf,Frederix:2015fyz,Hu:2021rkt} of the \minlo technique \cite{Hamilton:2012np}.
The \untwolops scheme has recently been generalised to processes with an additional jet in the context of an \unlops-based N$^3$LO+PS matching strategy \cite{Prestel:2021vww}. The \minlo-based schemes may be seen as a hybrid approach, since they use a combination of analytical and numerical resummation.
A noteworthy example of a scheme employing the latter approach is implemented in the \Geneva framework \cite{Alioli:2013hqa,Alioli:2021qbf}.
While all of these have enabled impressive phenomenological studies \cite{Hoche:2014uhw,Hoche:2014dla,Hoche:2018gti,Alioli:2015toa,Alioli:2016wqt,Cridge:2021hfr,Monni:2020nks,Lombardi:2020wju,Lombardi:2021rvg,Mazzitelli:2020jio,Buonocore:2021fnj} and provide pathways to matching precision calculations to event generators, they do not provide the same level of theoretical control as the fully-differential matching methods that are available at NLO.

In this letter, we present for the first time a fully-differential NNLO+PS matching scheme for final-state parton showers, restricting ourselves to the case of two coloured final-state particles. The new method combines NNLO antenna subtraction with the sector-antenna shower in \Vincia~\cite{Brooks:2020upa}, suitably extended to include real-virtual and double-real corrections.
A key aspect of the new technique is that the parton shower is employed only as an efficient Sudakov-weighted phase-space generator. It does not define the infrared subtraction terms that are key to \MCatNLO type matching strategies.

The letter is structured as follows. We review the matching method at NLO in \cref{sec:strategyNLO} before extending it to the NNLO in \cref{sec:strategyNNLO}, retaining a rather general notation. A numerical implementation in the \Vincia sector-antenna shower in \Pythia 8.3~\cite{Sjostrand:2014zea} is described in \cref{sec:implementation}, featuring a more detailed description of the matching scheme. We conclude in \cref{sec:conclusions} and provide an outlook on applications beyond the simple cases considered here.

\section{NLO Matching Strategy}\label{sec:strategyNLO}
Our matching strategy generalises the technique first developed in \cite{Norrbin:2000uu}, which nowadays is referred to as the \Powheg scheme \cite{Nason:2004rx,Frixione:2007vw,Alioli:2010xd}.
To start with, it is thus useful to recap the NLO matching strategy, before moving on to the new NNLO technique.

At NLO, the expected value of an infrared-safe observable $O$ defined on a two-particle final state process with a colourless initial state is given by
\begin{align}
   &\avg{O}_\mathrm{NLO} = \int \, \D \Phi_2 \left[ \UB(\Phi_2) + \UV(\Phi_2) + \UI^\nlo_{\US}(\Phi_2) \right] O(\Phi_2) \nonumber \\
   &+\int \, \D \Phi_3 \left[ \UR(\Phi_3) O(\Phi_3) - \US^\nlo(\Phi_3) O(\Phi_2(\Phi_3)) \right] \, ,
\label{eq:expvalNLO}
\end{align}
where $\UB$ and $\UV$ denote the Born cross section and virtual correction, differential in the two-particle phase space $\Phi_2$. Similarly, $\UR$ denotes the real-radiation cross section differential in the three-particle phase space $\Phi_3$, and $\US^\nlo$ denotes the differential NLO subtraction term in the antenna subtraction method, with its integral over the antenna phase space given by $\UI^\nlo_\US$.\footnote{Other well-established NLO subtraction schemes such as FKS~\cite{Frixione:1995ms} or dipole subtraction~\cite{Catani:1996vz} may equally well be employed here.}
In order to achieve a Born-local cancellation of the subtraction term upon integration over the real-emission phase space,
the observable acting on $\US^\nlo$ must be evaluated at the reduced phase-space point $\Phi_2(\Phi_3)$, where the precise mapping from the
three-parton to the two-parton state depends on the subtraction scheme.
We can invert this mapping and factorise the phase space into the 2-particle (Born) phase space $\Phi_2$, and the one-particle radiation phase space $\Phi_{+1}$,
\begin{equation}
    \D\Phi_{3} = \D\Phi_{2}\times\D\Phi_{+1} \, .
\end{equation}
By defining a Born-local NLO weight,
\begin{multline}\label{eq:local_kfac_nlo}
    k_\mathrm{NLO}(\Phi_2) :=  1 + \frac{\UV(\Phi_2)}{\UB(\Phi_2)} + \frac{\UI^\nlo_{\US}(\Phi_2)}{\UB(\Phi_2)} \\
    + \int \D \Phi_{+1}\, \left[\frac{\UR(\Phi_2,\Phi_{+1})}{\UB(\Phi_2)} - \frac{\US^\nlo(\Phi_2,\Phi_{+1})}{\UB(\Phi_2)}\right]
\end{multline}
\cref{eq:expvalNLO} can be rewritten as
\begin{multline}\label{eq:born_local_nlo}
    \avg{O}_\mathrm{NLO} = \int \D \Phi_2 \, \UB(\Phi_2)\, \Bigg[ k_\mathrm{NLO}(\Phi_2) O(\Phi_2)\\
    + \int \D \Phi_{+1}\,\frac{\UR(\Phi_2,\Phi_{+1})}{\UB(\Phi_2)} \left( O(\Phi_2,\Phi_{+1}) - O(\Phi_2) \right) \Bigg] \, .
\end{multline}

The parton-shower evolution, on the other hand, is described by a generating functional, the shower operator, recursively defined for an infrared-safe observable $O$ by
\begin{align}
    \CalS_n(t,O) &= \Delta_n(t,\tc) O(\Phi_n) \label{eq:showerOperator}\\
    & + \int^t_{\tc}\D\Phi_{+1}\, \UA^{(0)}_n(\Phi_{+1}) \Delta_n(t,t') \CalS_{n+1}(t',O) \, , \nonumber
\end{align}
where $\UA^{(0)}_n(\Phi_{+1})$ is the sum of all leading-order antenna functions\footnote{We refer to NLO antenna subtraction terms as LO antenna functions.} competing for the next branching $IK \mapsto ijk$ off the $n$-parton configuration,
\begin{align}\label{eq:ps_partitioning}
    &\int^t_{\tc}\D\Phi_{+1}\, \UA^{(0)}_{n\mapsto n+1}(\Phi_{+1}) \\
    &\qquad := \int^t_{\tc} \sum\limits_{j\in \{n\mapsto n+1\}}A_{j/IK}^{(0)}\left(\Phi_{+1}^j\right) \, \D\Phi_{+1}^j \, , \nonumber \\
    &\qquad = \sum\limits_{j\in \{n\mapsto n+1\}} \int^t_{\tc}\, \frac{\alphaS(t)}{4\uppi} \CalC_{j/IK} \bar{A}^{(0)}_{j/IK}(t,\zeta,\phi) \D t \D \zeta \frac{\D \phi}{2\uppi}\nonumber
\end{align}
with the sum and shower variables left implicit in our notation. 
Note that when working in the sector antenna framework~\cite{Kosower:1997zr,Kosower:2003bh,
 Larkoski:2009ah,Lopez-Villarejo:2011pwr,Brooks:2020upa}, 
Eq.~\eqref{eq:ps_partitioning} implicitly defines a partitioning
of the real-emission term along with the associated subtractions in Eq.~\eqref{eq:local_kfac_nlo}.
This is crucial to avoid double-counting of radiative corrections generated by the parton shower.
The associated Sudakov factor is given by
\begin{equation}
    \Delta_n^\mathrm{LO}(\tzero,t) = \exp{-\int^{\tzero}_{t}\D\Phi_{+1}\, \UA_{n\mapsto n+1}^{(0)}(\Phi_{+1})} \, .
\end{equation}

Taking only the first shower emission into account, the expected value of the observable $O$ at LO is given by
\begin{multline}
    \avg{O}_{\mathrm{LO}+\mathrm{PS}} = \int \D \Phi_2 \, \UB(\Phi_2)\, \Big[\Delta(\tzero,\tc) O(\Phi_2) \nonumber \\
    + \int \D \Phi_{+1}\, \UA^{(0)}_{2\mapsto3}(\Phi_{+1}) \Delta(\tzero,t) O(\Phi_2,\Phi_{+1}) \Big] \, .
\end{multline}
This implies that upon the replacement
\begin{align}
    \UB(\Phi_2) &\to k_\mathrm{NLO}(\Phi_2) \UB(\Phi_2) \nonumber\\
    \UA^{(0)}_{2\mapsto 3} &\to w^\mathrm{LO}_{2\mapsto3}(\Phi_2,\Phi_{+1}) \UA^{(0)}_{2\mapsto 3}
\end{align}
where we have defined the $2\mapsto 3$ LO matrix-element correction factor,
\begin{equation}
    w^\mathrm{LO}_{2\mapsto3}(\Phi_2,\Phi_{+1}) = \frac{\UR(\Phi_2,\Phi_{+1})}{\UA^{(0)}_{2\mapsto3}(\Phi_{+1})\UB(\Phi_2)} \, ,
\label{eq:LOMEC2to3}
\end{equation}
the following matching formula is NLO accurate up to terms appearing at order $\alphaS^2$
\begin{align}
    \avg{O}_{\mathrm{NLO}+\mathrm{PS}} = \int \D \Phi_2 \, \UB(\Phi_2)\, k_\mathrm{NLO}(\Phi_2)\Big[\Delta(\tzero,\tc) O(\Phi_2) \nonumber \\
    + \int^{\tzero}_{\tc} \D \Phi_{+1}\,  w^\mathrm{LO}_{2\mapsto3}\UA_{2\mapsto3}(\Phi_{+1}) \Delta(\tzero,t) O(\Phi_2,\Phi_{+1}) \Big] \, .
\end{align}
This can be seen by expanding the result to order $\alphaS$.

\section{NNLO Matching Strategy} \label{sec:strategyNNLO}
We now turn to the main result of this work, the definition of a fully-differential NNLO matching strategy for processes with two coloured final-state particles. It is applicable to both decays of colour singlets as well as scattering processes as long as all initial-state particles are colourless, for instance as in $e^+e^- \to jj$.

In the antenna formalism, the expected value for an infrared-safe observable of a process with two coloured final-state particles is given at NNLO by
\begin{align}
    \avg{O}_\mathrm{\small{NNLO}}
    &= \int \D\Phi_2 \, \Big[ \UB(\Phi_2) + \UV(\Phi_2) + \UI^\nlo_{\US}(\Phi_2) \nonumber \\
    &\qquad + \UVV(\Phi_2) + \UI_{\UT}(\Phi_2) + \UI_{\US}(\Phi_2)\Big]\, O(\Phi_2) \nonumber \\
    &+ \int \D\Phi_3 \,  \Big[ \UR(\Phi_2,\Phi_{+1})O(\Phi_2,\Phi_{+1}) \nonumber \\
    &\qquad - \US^\nlo(\Phi_2,\Phi_{+1})O(\Phi_2) \nonumber \\
    &\qquad + \URV(\Phi_2,\Phi_{+1})O(\Phi_2,\Phi_{+1}) \nonumber \\
    & \qquad + \UT(\Phi_2,\Phi_{+1},O) \Big] \nonumber \\
    &+ \int \D\Phi_4 \,  \Big[\URR(\Phi_2,\Phi_{+2})O(\Phi_2,\Phi_{+2}) \nonumber \\
    &\qquad - \US(\Phi_2,\Phi_{+2},O) \Big] \, ,
\label{eq:expvalNNLO}
\end{align}
where $\URR$ is the differential double-real radiation cross section and $\URV$ and $\UVV$ denote the differential virtual (one-loop) and double-virtual (two-loop) corrections to the real-radiation cross section $\UR$ and the Born cross section $\UB$, respectively.
In this context, the term $\US^\nlo$ denotes the differential NLO real antenna subtraction term, $\US$ denotes the differential NNLO double-real antenna subtraction term \cite{GehrmannDeRidder:2004tv,GehrmannDeRidder:2005aw,GehrmannDeRidder:2005hi,GehrmannDeRidder:2005cm},
\begin{multline}
    \US(\Phi_2,\Phi_{+2},O) = \US^{a}(\Phi_2,\Phi_{+2})O(\Phi_2) + \US^{b}(\Phi_3,\Phi_{+1})O(\Phi_3) \\
    - \US^c(\Phi_2,\Phi_{+1},\Phi_{+1}')O(\Phi_2)
\label{eq:SNNLORR}
\end{multline}
and $\UT$ denotes the differential NNLO real-virtual antenna subtraction term \cite{GehrmannDeRidder:2004tv,GehrmannDeRidder:2005aw,GehrmannDeRidder:2005hi,GehrmannDeRidder:2005cm},
\begin{multline}
    \UT(\Phi_2,\Phi_{+1},O) = \UT^{a}(\Phi_2,\Phi_{+1})O(\Phi_2) + \UT^{b}(\Phi_3)O(\Phi_3) \\
    - \UT^{c}(\Phi_2,\Phi_{+1})O(\Phi_2)
\label{eq:TNNLORV}
\end{multline}
Their integrated counterparts are given by $\UI^\nlo_{\US}$, $\UI_{\UT}$, and $\UI_{\US}$.
In this context, terms labelled with superscript $a$ constitute the double-real/real-virtual subtraction terms 
with compensating terms labelled with a superscript $c$ that remove spurious single-unresolved singularities.
The single-unresolved singularities are captured by the NLO subtraction terms of the $+1$-jet calculation, labelled with superscript $b$,
\begin{equation}
    \US^b(\Phi_3,\Phi_{+1}') \equiv {\US}^\nlo(\Phi_3,\Phi_{+1}')\, , 
    \quad \UT^b(\Phi_3) \equiv \UI_{\US}^\nlo(\Phi_3) \, .
\label{eq:subtTermsNLO1jet}
\end{equation}
The terms labeled with superscript $b$ and $c$ cancel independently in eq.~\eqref{eq:expvalNNLO}. 
They are constructed such as to make the integrals individually infrared finite and thus amenable
to evaluation with Monte-Carlo methods.

As for the NLO case, we define a Born-local weight,
\begin{align}\label{eq:born_local_nnlo_kfactor}
    k_\mathrm{NNLO}(\Phi_2) &:= 1 + \frac{\UV(\Phi_2)}{\UB(\Phi_2)} + \frac{\UI^\nlo_{\US}(\Phi_2)}{\UB(\Phi_2)} \\
    &+ \frac{\UVV(\Phi_2)}{\UB(\Phi_2)} + \frac{\UI_{\UT}(\Phi_2)}{\UB(\Phi_2)} + \frac{\UI_{\US}(\Phi_2)}{\UB(\Phi_2)} \nonumber \\
    &+ \int \D\Phi_{+1}\, \Big[\frac{\UR(\Phi_2,\Phi_{+1})}{\UB(\Phi_2)} - \frac{\US^\nlo(\Phi_2,\Phi_{+1})}{\UB(\Phi_2)} \nonumber \\
    &\qquad+\frac{\URV(\Phi_2,\Phi_{+1})}{\UB(\Phi_2)} + \frac{\UT(\Phi_2,\Phi_{+1})}{\UB(\Phi_2)} \Big] \nonumber \\
    &+ \int \D\Phi_{+2}\, \Big[\frac{\URR(\Phi_2,\Phi_{+2})}{\UB(\Phi_2)} - \frac{\US(\Phi_2,\Phi_{+2})}{\UB(\Phi_2)} \Big]\nonumber \, ,
\end{align}
which will be used to construct the NNLO matching formula, and which can be used to perform the fixed-order computation
in complete analogy to eq.~\eqref{eq:born_local_nlo}.
Here, $\D\Phi_{+2}$ is the two-particle radiation phase space that enters the factorised $n+2$-particle phase space
\begin{equation}
    \D\Phi_{n+2} = \D\Phi_{n} \times \D\Phi_{+2}\,.
\end{equation}
We shall further need to distinguish between an ordered and unordered component of the two-particle radiation phase space, according to the following partition of unity:
\begin{align}
    \D\Phi_{+2} &= \theta(t'-t)\D\Phi_{+2} + \theta(t-t')\D\Phi_{+2} \, ,\nonumber \\
    &= \D\Phi_{+2}^> + \D\Phi_{+2}^< \, .
\end{align}
The ordered part $\D\Phi_{+2}^<$ corresponds to the region accessible to strongly-ordered shower paths $\tzero > t > t'$, whereas the unordered part $\D\Phi_{+2}^>$ is inaccessible to strongly-ordered showers because of the larger intermediate scale $\tzero > t' > t$. We will use \Vincia's sector criterion, cf.\ sec.~3.3 in \cite{Brooks:2020upa}, to distinguish between the two, cf.~\cref{subsec:2to4}.

In order to be able to match the NNLO calculation with the shower, the shower needs to incorporate virtual corrections to ordinary $2\to 3$ branchings as well as new $2\to 4$ branchings, accounting for the simultaneous emission of two particles. These new shower terms correspond to the real-virtual and double-real corrections in the NNLO calculation. In addition, we need to incorporate the corresponding parton-shower counterterms.
We start by defining the two-particle NLO Sudakov as \cite{Li:2016yez}
\begin{multline}
    \Delta_2^\mathrm{NLO}(\tzero,t)  \\
    = \exp\Bigg\{-\int^{\tzero}_{t}\D\Phi_{+1}\, \UA_{2\mapsto3}^{(0)}(\Phi_{+1}) w^\mathrm{NLO}_{2\mapsto3}(\Phi_2,\Phi_{+1}) \Bigg\}  \\
    \times \exp\Bigg\{-\int^{\tzero}_{t}\D\Phi_{+2}^>\, \UA_{2\mapsto 4}^{(0)}(\Phi_{+2})w^\mathrm{LO}_{2\mapsto4}(\Phi_2,\Phi_{+2}) \Bigg\} \, ,
\label{eq:nloSudakov}
\end{multline}
where we have introduced the $2\mapsto4$ LO matrix-element correction factor,
\begin{equation}
    w^\mathrm{LO}_{2\mapsto4}(\Phi_2,\Phi_{+2}) = \frac{\URR(\Phi_2,\Phi_{+2})}{\UA^{(0)}_{2\mapsto4}(\Phi_{+2})\UB(\Phi_2)}
\label{eq:LOMEC2to4}
\end{equation}
and the $2\mapsto3$ NLO matrix-element correction factor $w^\mathrm{NLO}_{2\mapsto3}(\Phi_{+1})$, which we write in terms of a second order correction to the LO $2\mapsto 3$ MEC in \cref{eq:LOMEC2to3},
\begin{multline}
    w^\mathrm{NLO}_{2\mapsto3}(\Phi_2,\Phi_{+1}) = w^\mathrm{LO}_{2\mapsto3}(\Phi_2,\Phi_{+1})\\
    \times\big(1+\tilde{w}^\mathrm{FO}_{2\mapsto3}(\Phi_2,\Phi_{+1})
    +\tilde{w}^\mathrm{PS}_{2\mapsto3}(\Phi_2)\big)\,.
\label{eq:NLOMEC2to3}
\end{multline}
The coefficients $\tilde{w}$ are given by matching the $\Order{\alphaS^2}$ terms
in the expansion of the truncated shower approximation to the fixed-order result 
in \cref{eq:expvalNNLO} \cite{Li:2016yez,Hartgring:2013jma}.
We find the fixed-order contribution
\begin{multline}
    \tilde{w}^\mathrm{FO}_{2\mapsto3}(\Phi_2,\Phi_{+1})=\\
    \frac{\URV(\Phi_2,\Phi_{+1})}{\UR(\Phi_2,\Phi_{+1})} + \int^{t}_{0}\D\Phi_{+1}^\prime\, \frac{\URR(\Phi_2,\Phi_{+1},\Phi_{+1}^\prime)}{\UR(\Phi_2,\Phi_{+1})} \\
    -\left(\frac{\UV(\Phi_2)}{\UB(\Phi_2)}+\int^{\tzero}_{0}\D\Phi_{+1}'\,\frac{\UR(\Phi_2,\Phi_{+1}')}{\UB(\Phi_2)}\right)\;,
\label{eq:VMEC2to3}
\end{multline}
and the second-order parton-shower matching term
\begin{multline}
    \tilde{w}^\mathrm{PS}_{2\mapsto3}(\Phi_2)
    = \frac{\alphaS}{2\pi}\ln\frac{\kappa^2\mu_{\rm S}^2}{\mursq}\\
    + \int^{\tzero}_{t}\D\Phi_{+1}'\, \UA_{2\mapsto3}^{(0)}(\Phi_{+1}')w^\mathrm{LO}_{2\mapsto3}(\Phi_2,\Phi_{+1}')\;.
\end{multline}
The factor $\kappa$ is a constant and $\mu_\mathrm{S}^2$ is the parton-shower renormalisation scale.
The two are conventionally chosen such that the logarithmic structure of eq.~\eqref{eq:VMEC2to3}
is reproduced, which leads to $\mu_\mathrm{S}=p_\perp$ and $\kappa^2=\exp\{K/\beta_0\}$, with $K$ 
the two-loop cusp anomalous dimension~\cite{Kodaira:1981nh,Davies:1984hs,Davies:1984sp,Catani:1988vd}. 
This is known as the CMW scheme~\cite{Catani:1990rr}.

Note that in \cref{eq:nloSudakov}, the integral over $\UA_{2\mapsto 4}^{(0)}$ is defined over 
the range $[t, \tzero]$, since the ``ordered'' contribution $t'<t$ has been reabsorbed 
into $\tilde w^\mathrm{FO}_{2\mapsto3}(\Phi_{+1})$.

It should be emphasised that we do not require the NLO three-jet calculation to be provided externally 
but include the correction directly in the shower evolution. This means that, different to 
the situation in traditional merging approaches, this correction is exponentiated into a Sudakov factor.
Up to the first emission, this agrees with the treatment in~\cite{Hoche:2017iem,Dulat:2018vuy}
and implicitly includes the contribution from higher-order matching terms and collinear mass 
factorization counterterms that are needed to recover the NLO DGLAP splitting functions.

In addition, we need the $3$-particle Sudakov, which we describe at LO,
\begin{multline}
    \Delta_3^\mathrm{LO}(t,t') \\
    = \exp\Bigg\{-\int^{t}_{t'}\D\Phi^\prime_{+1}\, \UA_{3\mapsto4}^{(0)}(\Phi^\prime_{+1}) w^\mathrm{LO}_{3\mapsto4}(\Phi_3,\Phi_{+1}^\prime) \Bigg\} \, .
\end{multline}
with the $3\mapsto 4$ LO matrix-element correction factor,
\begin{equation}
    w^\mathrm{LO}_{3\mapsto4}(\Phi_3,\Phi_{+1}^\prime) 
    = \frac{\URR(\Phi_2,\Phi_{+2})}{\UA^{(0)}_{3\mapsto4}(\Phi_{+1}^\prime)\UR(\Phi_2,\Phi_{+1})} \, .
\label{eq:LOMEC3to4}
\end{equation}

Up to the second emission, the shower operator is thus given by
\begin{align}
    &\CalS_2(\tzero,O) = \Delta_2^\mathrm{NLO}(\tzero,\tc)O(\Phi_2) \label{eq:showerOpNLO} \\
    &+ \int^{\tzero}_{\tc}\D\Phi_{+1}\, \UA_{2\mapsto3}^{(0)}(\Phi_{+1})w^\mathrm{NLO}_{2\mapsto3}\Delta^\mathrm{NLO}_2(\tzero,t) \nonumber \\
    &\quad \times \Big(\Delta_3^\mathrm{LO}(t,\tc)O(\Phi_2,\Phi_{+1})\nonumber \\
    &\qquad + \int^t_{\tc}\D\Phi_{+1}^\prime\, \UA^{(0)}_{3\mapsto4}(\Phi^\prime_{+1})  w^\mathrm{LO}_{3\mapsto4}(\Phi_3,\Phi_{+1}^\prime)O(\Phi_3,\Phi_{+1}^\prime) \Big) \nonumber \\
    &+ \int^{\tzero}_{\tc}\D\Phi_{+2}^>\, \UA_{2\mapsto4}^{(0)}(\Phi_{+2})w^\mathrm{LO}_{2\mapsto4} (\Phi_2,\Phi_{+2})O(\Phi_2,\Phi_{+2}) \nonumber
\end{align}
and our final NNLO+PS matching formula takes the simple form:
\begin{equation}
    \avg{O}_{\mathrm{NNLO}+\mathrm{PS}} = \int \D \Phi_2\, \UB(\Phi_2) k_\mathrm{NNLO}(\Phi_2) \CalS_2(\tzero,O) \, .
\label{eq:nnlops}
\end{equation}

When expanding the truncated shower operator $\CalS_2$ in \cref{eq:nnlops} up to order $\alphaS^2$, NNLO accuracy is recovered for the observable $O(\Phi_2)$, while $O(\Phi_3)$ and $O(\Phi_4)$ achieve NLO and LO accuracy, respectively.
This is true, because the combination of the iterated $2\mapsto 3\mapsto 4$ and the direct $2\mapsto 4$ contributions to \cref{eq:showerOpNLO} yields the correct double-real correction $\URR$ in \cref{eq:expvalNNLO} by means of the LO MEC factors in \cref{eq:LOMEC2to3,eq:LOMEC3to4,eq:LOMEC2to4}.  Moreover the NLO correction \cref{eq:NLOMEC2to3} recovers the correct real and real-virtual corrections $\UR$ and $\URV$ in \cref{eq:expvalNNLO} by means of \cref{eq:LOMEC2to3} and \cref{eq:VMEC2to3}.

\section{Numerical Implementation} \label{sec:implementation}
In this section, we want to present all necessary components of an implementation of our NNLO matching strategy. These are:
\begin{itemize}
    \item a framework to calculate the Born-local NNLO $K$-factors in Eq.~\eqref{eq:born_local_nnlo_kfactor}
    \item a shower filling the strongly-ordered \cite{Brooks:2020bhi} and unordered \cite{Li:2016yez} regions of the single- and double-emission phase space
    \item tree-level MECs in strongly-ordered \cite{Fischer:2017yja} and unordered \cite{Giele:2011cb} shower paths
    \item NLO MECs in the first emission \cite{Hartgring:2013jma}
\end{itemize}
With the exception of the first point, (process-dependent) implementations of these components existed in previous \Vincia versions (not necessarily simultaneously), and have been described in detail in the various references.
We have (re-)implemented all components in a semi-automated~\footnote{Semi-automated here refers to the fact that antenna subtraction terms are explicitly implemented for each class of processes.} fashion in the \Vincia antenna shower in \Pythia 8.3. We access loop matrix elements via a novel \MCFM \cite{Campbell:1999ah,Campbell:2011bn,Campbell:2015qma,Campbell:2019dru} interface presented in \cite{Campbell:2021vlt} and tree-level matrix elements via a new run-time interface \cite{ComixInterface} to the \Comix matrix element generator \cite{Gleisberg:2008fv} in \Sherpa \cite{Gleisberg:2008ta,Sherpa:2019gpd}.

Our NNLO matching algorithm can be summarised in the following steps:
\begin{enumerate}
    \item[1.] Generate a phase space point according to the Born cross section $\UB(\Phi_2)$.
    \item[2.] Calculate the Born-local NNLO factor $k_\mathrm{NNLO}(\Phi_2)$ and reweight the phase space point by the result. 
    \item[3.] Let the phase-space maximum given by the invariant mass of the two Born partons define the starting scale for the shower, $t_\mathrm{now} = t_0(\Phi_2)$.
    \item[4.] Starting from the current shower scale, $t_\mathrm{now}$, let the $2\mapsto 3$ and $2\mapsto 4$ showers compete for the highest branching scale. 
    \item[5.] Update the current shower scale to be that of the winning branching, $t_\mathrm{now} = \mathrm{max}(t_{2\mapsto 3},t_{2\mapsto 4})$.
    \item[6a.] If the winning branching is a $2\mapsto 3$ branching, calculate the accept probability including the NLO MEC $w^\mathrm{NLO}_{2\mapsto3}$.  
    \begin{itemize}
        \item  If rejected, continue from step 4.
        \item  If accepted, continue with a LO shower from the resulting three-particle configuration, starting from $t_\mathrm{now}$ and including the LO MEC $w^\mathrm{LO}_{3\mapsto 4}$ when calculating accept probabilities for the $3\mapsto4$ step. 
    \end{itemize}
    When a $3\mapsto 4$ branching is accepted (or the shower cutoff scale is reached), continue with step 7.
    \item[6b.] If the winning branching is a $2\mapsto 4$ branching, calculate the accept probability including the LO MEC $w^\mathrm{LO}_{2\mapsto4}$. 
    \begin{itemize}
        \item If rejected, continue from step 4. 
        \item If accepted, continue with step 7.
    \end{itemize}
    \item[7.] Continue with a standard (possibly uncorrected) shower from the resulting four-particle configuration, starting from $t_\mathrm{now}$. 
\end{enumerate}
It should be emphasised that the matrix-element correction factors make this algorithm independent of the splitting kernels (i.e.\ antenna functions in our case) up to the matched order and the shower merely acts as an efficient Sudakov-weighted phase-space generator. Hence, if the algorithm is stopped after step 6, an NNLO-matched result is obtained, which can be showered by any other parton shower, just as is the case for \Powheg NLO matching. Note, that there remains a dependence on the ordering variable, which has to be properly accounted for.

\subsection{NNLO Kinematics}\label{subsec:kinematics}
For both, the unordered shower contributions and the Born-local NNLO weight, new kinematic maps are needed to reflect their direct $2\mapsto 4$, i.e.\ unordered or double-unresolved, nature. We utilise that the $n$-particle phase space measure
may be factorised into the product of a $2\mapsto 3$ antenna phase space and the $n-1$-particle phase space measure, as well as into the product of a $2\mapsto 4$ antenna phase space and the $n-2$-particle phase space.
This allows us to write the $2\mapsto 4$ antenna phase space as the product of two $2\mapsto 3$ antenna phase spaces,
\begin{multline}
    \D \Phi_{+2} (p_I+p_K; p_i, p_{j_1}, p_{j_2}, p_{k}) \\
    = \D \Phi_{+1} (p_I+p_K; \hat{p}_i, \hat{p}_{j}, p_{k}) \\
    \times \D \Phi_{+1} (\hat{p}_i+\hat{p}_j; p_i, p_{j_1}, p_{j_2}) \, ,
\label{eq:PS2to4}
\end{multline}
corresponding to the kinematic mapping
\begin{equation}
    p_I + p_K = \hat{p}_i + \hat{p}_j + p_k = p_i + p_{j_1} + p_{j_2} + p_k \, ,
\end{equation}
effectively representing a tripole map \cite{Gehrmann-DeRidder:2003pne}. In line with the phase space factorisation, the kinematic mapping is then constructed as an iteration of two on-shell $2\mapsto 3$ antenna maps given in sec.~2.3 in \cite{Brooks:2020upa}. 

We have tested the validity of our kinematic maps by comparing \Vincia's phase-space mappings (double-gluon emission and gluon-emission-plus-splitting) to a flat sampling via \Rambo.

\subsection{Unordered Shower Contributions}\label{subsec:2to4}
\begin{figure*}[t]
    \centering
    \includegraphics[width=0.4\textwidth]{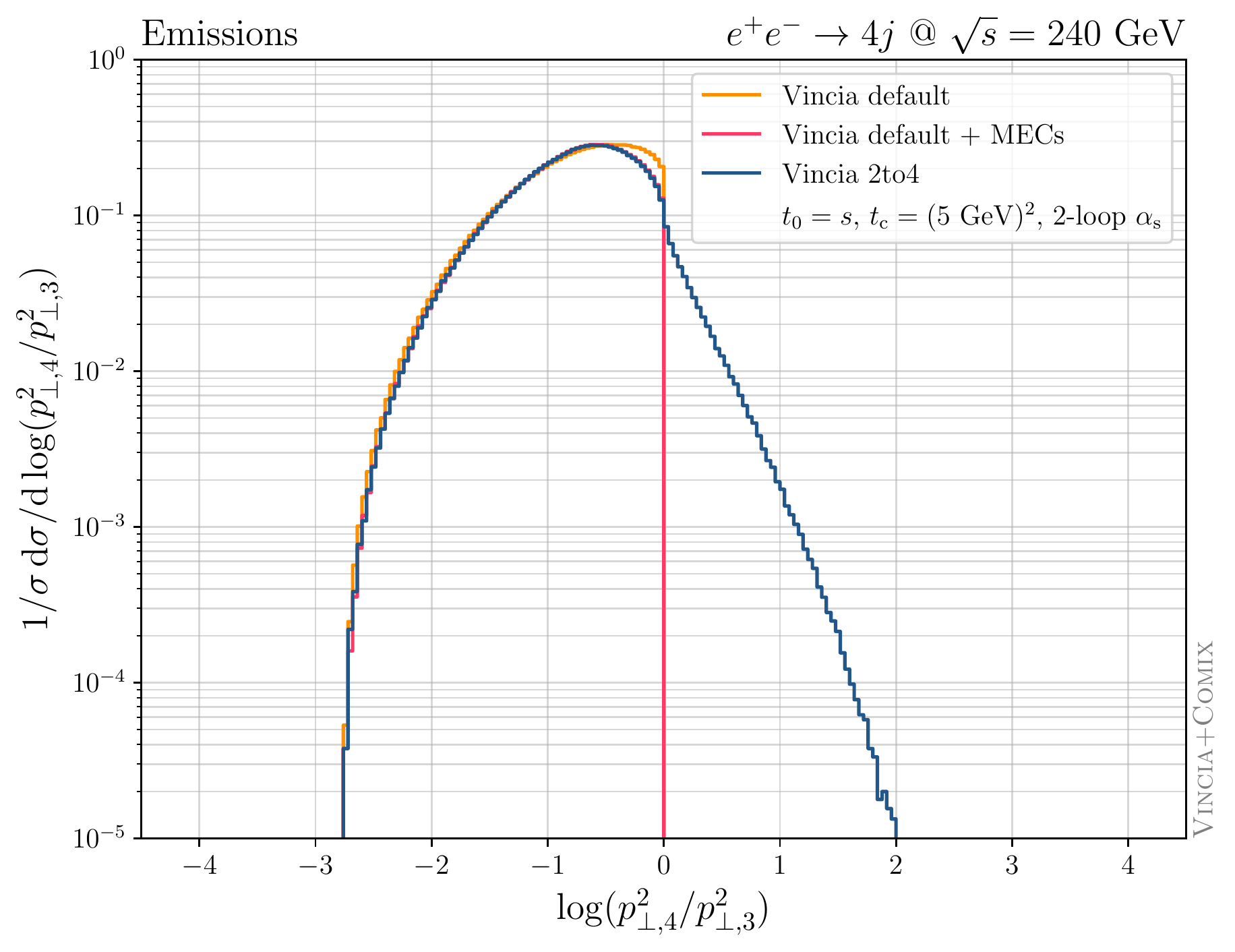}
    \includegraphics[width=0.4\textwidth]{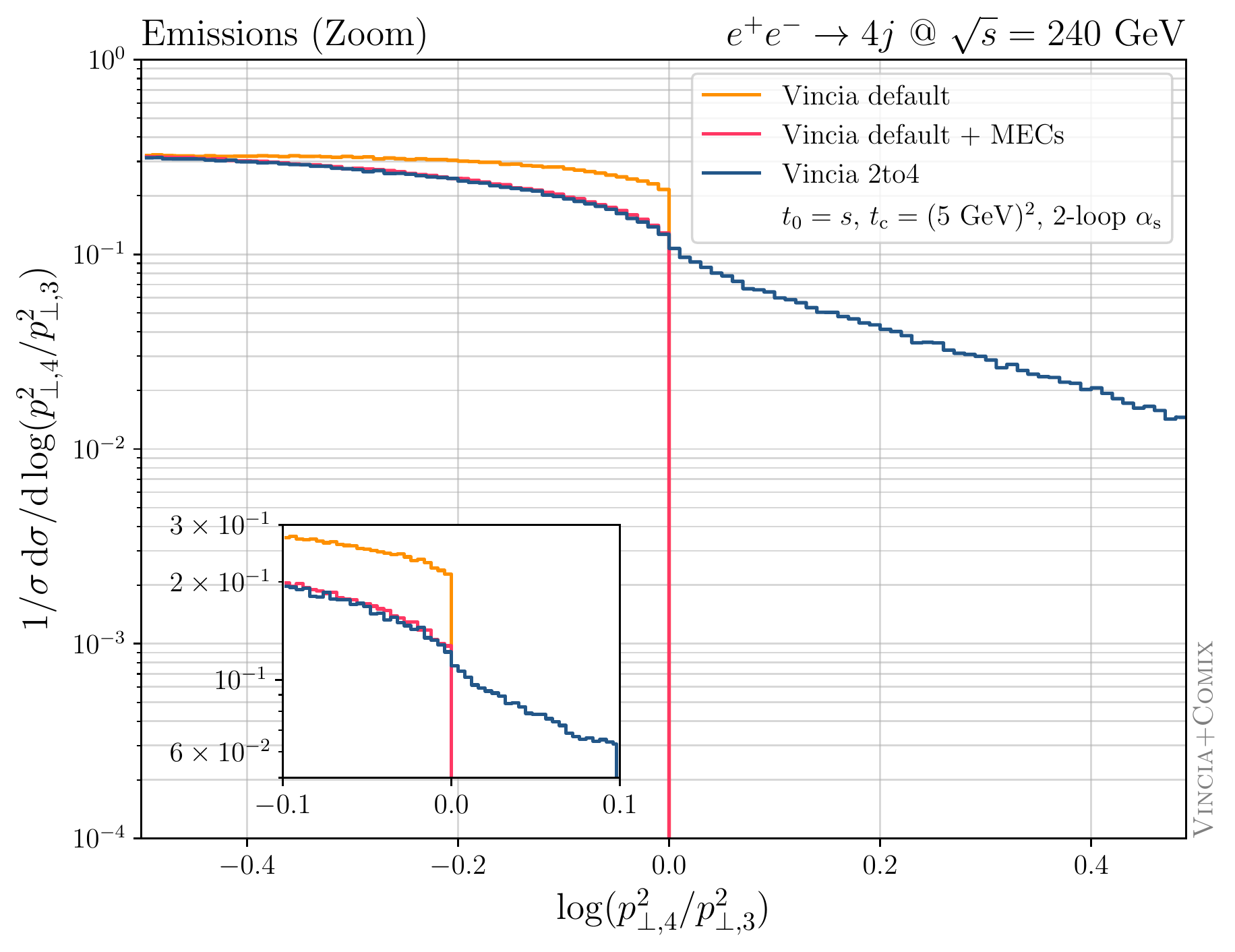}
    \includegraphics[width=0.4\textwidth]{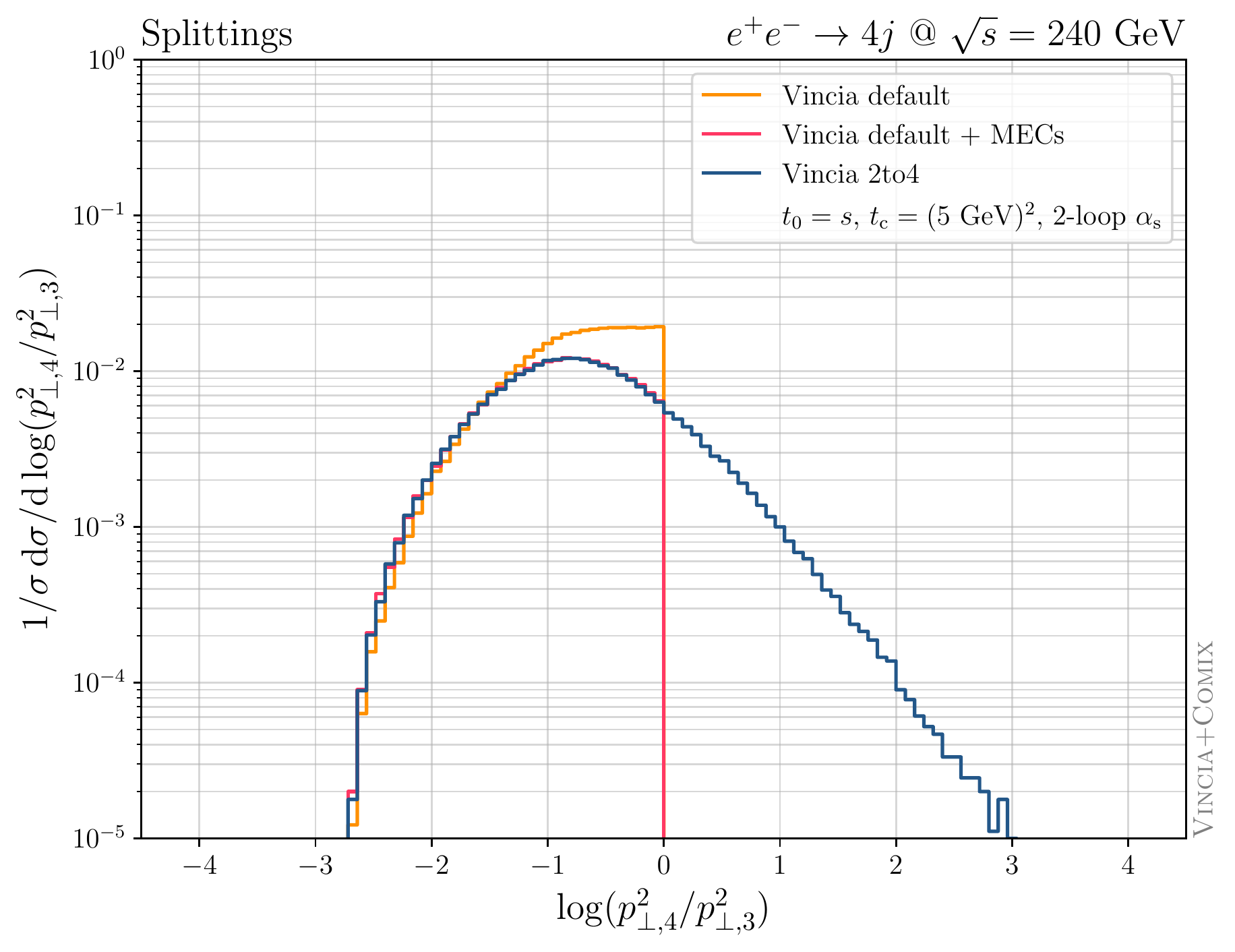}
    \includegraphics[width=0.4\textwidth]{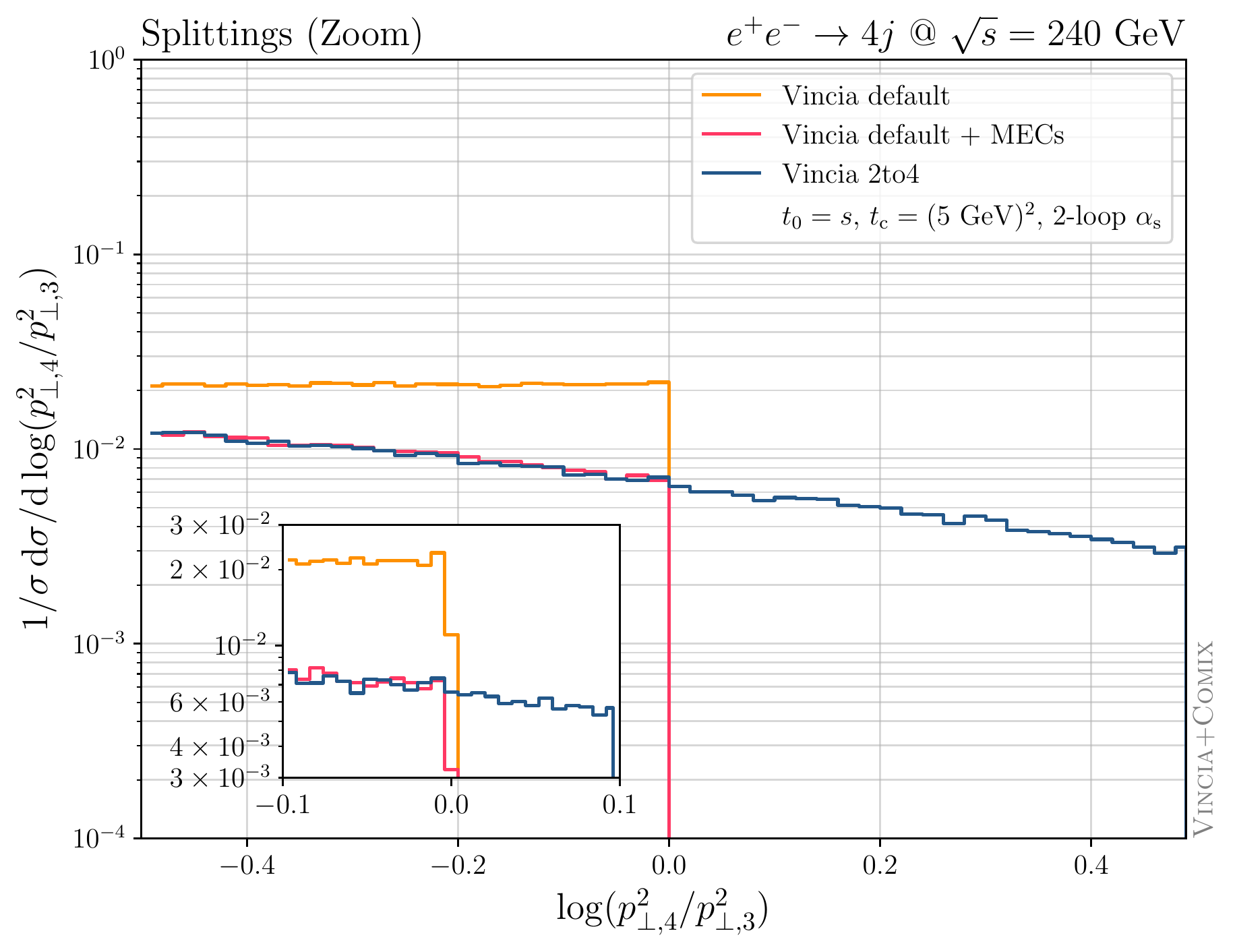}
    \caption{Ratio of the evolution variable of the four-parton and three-parton configuration $\log(p_{\perp,4}^2/p_{\perp,3}^2)$ in $e^+e^-\to 4j$. The region $> 0$ corresponds to unordered contributions not reached by strongly-ordered showers.}
    \label{fig:orderingZDecays}
\end{figure*}

An important part of our proposal is the inclusion of double-unresolved radiation in the shower evolution.
To this end, we employ the sector-antenna framework \cite{Brooks:2020upa} and amend it by direct $2\mapsto 4$ branchings as described in \cite{Li:2016yez}. 
In the sector-shower approach, each branching is restricted to the region in phase space where it minimises the resolution variable, defined for final-state clusterings by
\begin{equation}
    Q^2_{\mathrm{res},j} = \begin{cases} \frac{s_{ij}s_{jk}}{s_{IK}} & \text{if } j \text{ is a gluon} \\ s_{ij} \sqrt{\frac{s_{jk}}{s_{IK}}} & \text{if } (i,j) \text{ is a quark-antiquark pair} \end{cases}
\label{eq:resVar}
\end{equation}
This is achieved by a ``sectorisation'' of phase space according the partition of unity,
\begin{equation}
    1 = \sum\limits_j \Theta^\mathrm{sct}_{j/IK} = \sum\limits_j \theta\left(\min\limits_{i}\left\{Q^2_{\mathrm{res},i}\right\} - Q^2_{\mathrm{res},j}\right) \, ,
\label{eq:sectorVetoLO}
\end{equation}
which is implemented in the shower evolution as an explicit veto for each trial branching.
Since only a single branching kernel contributes per colour-ordered phase space point, sector antenna functions have to incorporate the full singularity structure associated with the respective sector. At LO, this amounts to including both the full single-collinear and single-soft limits in the antenna function. The full set of \Vincia's LO sector antenna functions is collected in \cite{Brooks:2020upa}.

By construction, the default sector shower generates only strongly-ordered sequences\footnote{This is different to virtually any other strongly-ordered shower, where recoil effects introduce unordered sequences. Such phase space points are vetoed in a sector shower.}, as the sector veto ensures that each emission is the softest (or most-collinear) in the post-branching configuration. The inclusion of direct $2\mapsto 4$ branchings (which look unordered from an iterated $2\mapsto 3$ point of view) in the sector shower is facilitated by extending the sector decomposition in \cref{eq:sectorVetoLO} by an ordering criterion,
\begin{align}
    1 &= \sum\limits_j \left[\Theta^<_{j/IK}\Theta^\mathrm{sct}_{j/IK} + \Theta^>_{j/IK}\Theta^\mathrm{sct}_{j/IK}\right]\\
    &= \underbrace{\sum\limits_j \theta\left(\hat{p}_{\perp,\hat{j}}^2 - p_{\perp,j}^2\right)\Theta^\mathrm{sct}_{j/IK}}_{2\mapsto 3 \text{ (strongly ordered)}} + \underbrace{\sum\limits_j \theta\left(p_{\perp,j}^2 - \hat{p}_{\perp,\hat{j}}^2\right)\Theta^\mathrm{sct}_{j/IK}}_{2\mapsto4 \text{ (unordered)}} \nonumber 
\end{align}
where $p_\perp^2$ denotes \Vincia's transverse-momentum ordering variable and hatted variables denote the intermediate node in a sequence $IL \mapsto \hat{i} \hat{j} \hat{\ell} \mapsto i j k \ell$. Here, the scales $p_\perp^2$ and $\hat p_\perp^2$ are uniquely defined by the ordering variable of the sector-shower emission, i.e., that emission which minimises \cref{eq:resVar}. Direct $2\mapsto 4$ emissions are thus restricted to the unordered region of the double-emission phase space, denoted as $\D\Phi_{+2}^>$ in \cref{eq:nloSudakov} and defined as
\begin{equation}
    \D\Phi_{+2}^> = \sum\limits_{j} \Theta^>_{j/IK} \Theta^\mathrm{sct}_{j/IK} \D \Phi_{+2}^j \, .
\end{equation}

For $2\to 4$ emissions off quark-antiquark and gluon-gluon antennae, we use the double-real antenna functions in \cite{GehrmannDeRidder:2004tv,GehrmannDeRidder:2005aw,GehrmannDeRidder:2005cm}. We note that NLO quark-gluon antenna functions appear in the Standard Model at lowest order for three final-state particles and are hence not of interest for our test case of $e^+e^-\to jj$. We wish to point out, however, that the NLO quark-gluon antenna functions in \cite{GehrmannDeRidder:2005hi,GehrmannDeRidder:2005cm} contain spurious singularities which have to be removed before a shower implementation is possible.

As a validation, we show in \cref{fig:orderingZDecays} the ratio of the four-jet to three-jet evolution variable for $e^+e^- \to 4j$ at $\sqrt{s} = 240~\giga e\volt$. To focus on the perturbative realm, the shower evolution is constrained to the region between $\tzero = s$ and $\tc = (5~\giga e\volt)^2$. The region $>0$ corresponds to the unordered part of phase space to which strongly-ordered showers cannot contribute. Due to the use of sector showers, there is a sharp cut-off at the boundary between the ordered and unordered region, as the sector criterion ensures that the last emission is always the softest and therefore, no recoil effects can spoil the strong ordering of the shower. As expected, the inclusion of direct $2\to 4$ branchings gives access to the unordered parts of phase space, a crucial element of our matching method.

\subsection{LO Matrix-Element Corrections}
In order for the shower expansion to match the fixed-order calculation, we need (iterated) $2\mapsto 3$ tree-level MECs and (direct) $2\mapsto4$ tree-level MECs. Both take a particularly simple form in the sector-antenna framework, as will be shown below.

At leading-colour, tree-level MECs to the ordered sector shower can be constructed as \cite{LopezVillarejo:2011ap,Fischer:2017yja}
\begin{align*}
    w_{2\mapsto 3,i}^{\mathrm{LO},\mathrm{LC}}(\Phi_2,\Phi_{+1})
    &= \frac{\UR^\mathrm{LC}_{i}(\Phi_2,\Phi_{+1})}{\sum_j \Theta^\mathrm{sct}_{j/IK} A^\mathrm{sct}_{j/IK}(p_i,p_j,p_k)\UB(\Phi_2)} \, ,\\
    w_{3\mapsto 4,i}^{\mathrm{LO},\mathrm{LC}}(\Phi_3,\Phi_{+1})
    &= \frac{\URR^\mathrm{LC}_{i}(\Phi_3,\Phi_{+1})}{\sum_j \Theta^\mathrm{sct}_{j/IK} A^\mathrm{sct}_{j/IK}(p_i,p_j,p_k)\UR^\mathrm{LC}_i(\Phi_3)} \, ,
\end{align*}
where
\begin{align*}
    \UB(\Phi_2) &= \abs{\CalM_2^{(0)}(p_1,p_2)}^2 \, , \\
    \UR^\mathrm{LC}_{i}(\Phi_3) &= \abs{\CalM_3^{(0)}(\sigma_i\{p_1,p_2,p_3\})}^2 \, , \\
    \URR^\mathrm{LC}_{i}(\Phi_4) &= \abs{\CalM_4^{(0)}(\sigma_i\{p_1,p_2,p_3,p_4\})}^2 \, ,
\end{align*}
denote squared leading-colour colour-ordered amplitudes with the index $i$ denoting the respective permutation $\sigma_i$ (the number of permutations depends on the process). The sector veto $\Theta^\mathrm{sct}_{j/IK}$ ensures that only the most singular term contributes in the denominators, rendering the fraction exceptionally simple.

Direct $2\mapsto 4$ branchings can be corrected in an analogous way, replacing the sum over $2\mapsto3$ antenna functions with a sum of $2\mapsto4$ ones,
\begin{multline*}
    w_{2\mapsto 4,i}^{\mathrm{LO},\mathrm{LC}}(\Phi_2,\Phi_{+2})
    \\
    = \frac{\URR^\mathrm{LC}_{i}(\Phi_2,\Phi_{+2})}{\sum_{\{j,k\}} \Theta^\mathrm{sct}_{jk/IL} A^\mathrm{sct}_{jk/IL}(p_i,p_j,p_k,p_\ell)\UB(\Phi_2)} \, ,
\end{multline*}

The full-colour matrix element can be recovered on average by multiplication with a full-colour to leading-colour-summed matrix-element weight,
\begin{align}
    w_{2\mapsto 3,i}^{\mathrm{LO}} &= w_{2\mapsto 3,i}^{\mathrm{LO},\mathrm{LC}}
    \times \frac{\UR(\Phi_2,\Phi_{+1})}{\sum_j \UR^\mathrm{LC}_{j}(\Phi_2,\Phi_{+1})} \, , \\
    w_{3\mapsto 4,i}^{\mathrm{LO}} &= w_{3\mapsto 4,i}^{\mathrm{LO},\mathrm{LC}}
    \times \frac{\URR(\Phi_3,\Phi_{+1})}{\sum_j \URR^\mathrm{LC}_{j}(\Phi_3,\Phi_{+1})} \, , \\
    w_{2\mapsto 4,i}^{\mathrm{LO}} &= w_{2\mapsto 4,i}^{\mathrm{LO},\mathrm{LC}}
    \times \frac{\URR(\Phi_2,\Phi_{+2})}{\sum_j \URR^\mathrm{LC}_{j}(\Phi_2,\Phi_{+2})} \, .
\end{align}

For gluon splittings, multiple histories contribute even in the sector shower, because all permutations of quark lines have to be taken into account. To ensure that the MEC factors remain finite for final states with multiple quark pairs, an additional quark-projection factor has to be included. Since we only deal with a maximum of two quark pairs, it is given by
\begin{equation}
    \rho_j = \frac{A^\mathrm{sct}_{j_q/g_IX_K}(\bar q_i, q_j, X_k)}{\sum_{j}A^\mathrm{sct}_{j_q/g_IX_K}(\bar q_i, q_j, X_k)}
\end{equation}
for $2\to 3$ branchings and
\begin{equation}
    \rho_j = \frac{A^\mathrm{sct}_{j_qk_{\bar q}/X_IY_L}(X_i, q_j,\bar q_k, Y_\ell)}{\sum_{j}A^\mathrm{sct}_{j_qk_{\bar q}/X_IY_L}(X_i, q_j, \bar q_k, Y_\ell)}
\end{equation}
for $2\mapsto 4$ branchings.

\subsection{NLO Matrix-Element Corrections}
Making the antenna subtraction terms explicit, the fixed-order correction to the NLO matrix-element correction \cref{eq:NLOMEC2to3} reads
\begin{align}
    &\tilde{w}^\mathrm{FO}_{2\mapsto3}(\Phi_2,\Phi_{+1}) = \frac{\URV(\Phi_2,\Phi_{+1})}{\UR(\Phi_2,\Phi_{+1})} + \frac{\UI^\nlo(\Phi_2,\Phi_{+1})}{\UR(\Phi_2,\Phi_{+1})}
    \label{eq:VMEC2to3Numerical} \\
    &\, + \int^{t}_{0}\D\Phi_{+1}^\prime\, \left[ \frac{\URR(\Phi_2,\Phi_{+1},\Phi_{+1}^\prime)}{\UR(\Phi_2,\Phi_{+1})} - \frac{\US^\nlo(\Phi_2,\Phi_{+1},\Phi_{+1}^\prime)}{\UR(\Phi_2,\Phi_{+1})} \right] \nonumber\\
    &\, - \Bigg(\frac{\UV(\Phi_2)}{\UB(\Phi_2)} + \frac{\UI^\nlo(\Phi_2)}{\UB(\Phi_2)} \nonumber \\
    &\qquad + \int^{\tzero}_{0}\D\Phi_{+1}'\, \left[\frac{\UR(\Phi_2,\Phi_{+1}^\prime)}{\UB(\Phi_2)} - \frac{\US^\nlo(\Phi_2,\Phi_{+1}^\prime)}{\UB(\Phi_2)} \right]\Bigg)\, , \nonumber
\end{align}
with the differential NLO antenna subtraction terms $\US^\nlo(\Phi_2,\Phi_{+1}^\prime)$, $\US^\nlo(\Phi_2,\Phi_{+1},\Phi_{+1}^\prime)$ and their integrated counterparts $\UI^\nlo_{\US}(\Phi_2)$, $\UI^\nlo_{\US}(\Phi_2,\Phi_{+1})$ cf.~\cref{eq:expvalNNLO,eq:subtTermsNLO1jet}.
Based on the argument of the last subsection, we construct the full-colour NLO matrix-element correction as
\begin{align}
   w^\nlo_{2\mapsto 3,i}(\Phi_2,\Phi_{+1}) &= w^{\mathrm{LO},\mathrm{LC}}_{2\mapsto3,i}(\Phi_2,\Phi_{+1}) \frac{\UR(\Phi_2,\Phi_{+1})}{\sum_j \UR^\mathrm{LC}_{j}(\Phi_2,\Phi_{+1})} \nonumber \\ 
   &\quad \times (1 + \tilde{w}^\text{FO}_{2\mapsto 3}(\Phi_2,\Phi_{+1}) + \tilde{w}^\text{PS}_{2\mapsto 3}(\Phi_2)) \, .
\end{align}

The integration over the radiation phase spaces denoted $\Phi_{+1}^\prime$ in \cref{eq:VMEC2to3Numerical} is done numerically, utilising antenna kinematics to map $3$-parton configurations to $4$-parton configurations (similarly for $2$-parton configurations). This phase-space generation approach will be described in detail in the next subsection in the context of the NNLO Born weight.
Note that the radiation phase space $\Phi_{+1}$ in \cref{eq:VMEC2to3Numerical} is generated by the shower.

\subsection{NNLO Born Weight}
The Born-local NNLO weight can be calculated numerically using a ``forward-branching'' phase-space generation approach \cite{Frixione:2007vw,Hoche:2010pf,Alioli:2010xd,Giele:2011tm,Figy:2018imt}, which has previously been applied to unweighted NLO event generation, using Catani-Seymour dipole subtraction \cite{Campbell:2012cz}. The application to NNLO corrections to $e^+e^- \to 2j$ using antenna subtraction has been outlined in \cite{Weinzierl:2006ij}.

Given a Born phase space point, the real-radiation phase space is generated by uniformly sampling the shower variables $(t,\zeta,\phi)$ for each antenna, which represent integration channels in this context. As for the shower evolution, every phase space point is restricted to the sector in which the emission(s) correspond to the most-singular clusterings. 
The momenta of the Born$+1j$ point are constructed according to the same kinematic map as the shower uses, summarised in sec.~2.3 in \cite{Brooks:2020bhi}. Since antenna functions are azimuthally averaged, they do not cancel spin-correlations in collinear gluon branchings locally. To obtain a point-wise pole cancellation, the subtracted real correction $\UR-\US$ can be evaluated on two correlated phase space points,
\begin{equation*}
    \left\{ \left(t,\zeta,\phi\right), \left(t,\zeta,\phi+\uppi/2\right)\right\} \,
\end{equation*}
which cancels the collinear spin correlation exactly, as it is proportional to $\cos(2\phi)$.
To obtain double-real radiation phase space points for the subtracted double-real correction $\URR-\US$, this procedure can be iterated, yielding four angular-correlated phase space points which cancel spin correlations in double single-collinear and triple-collinear limits.
Due to the bijective nature of the sector-antenna framework, each $3$- or $4$-particle phase-space point obtained in this way can be mapped back uniquely to its $2$-particle origin, making the NNLO weight exactly Born-local. For $e^+e^- \to 2j$ this procedure is identical to the one in \cite{Weinzierl:2006ij}.

\begin{figure*}[t]
    \centering
    \includegraphics[width=0.4\textwidth]{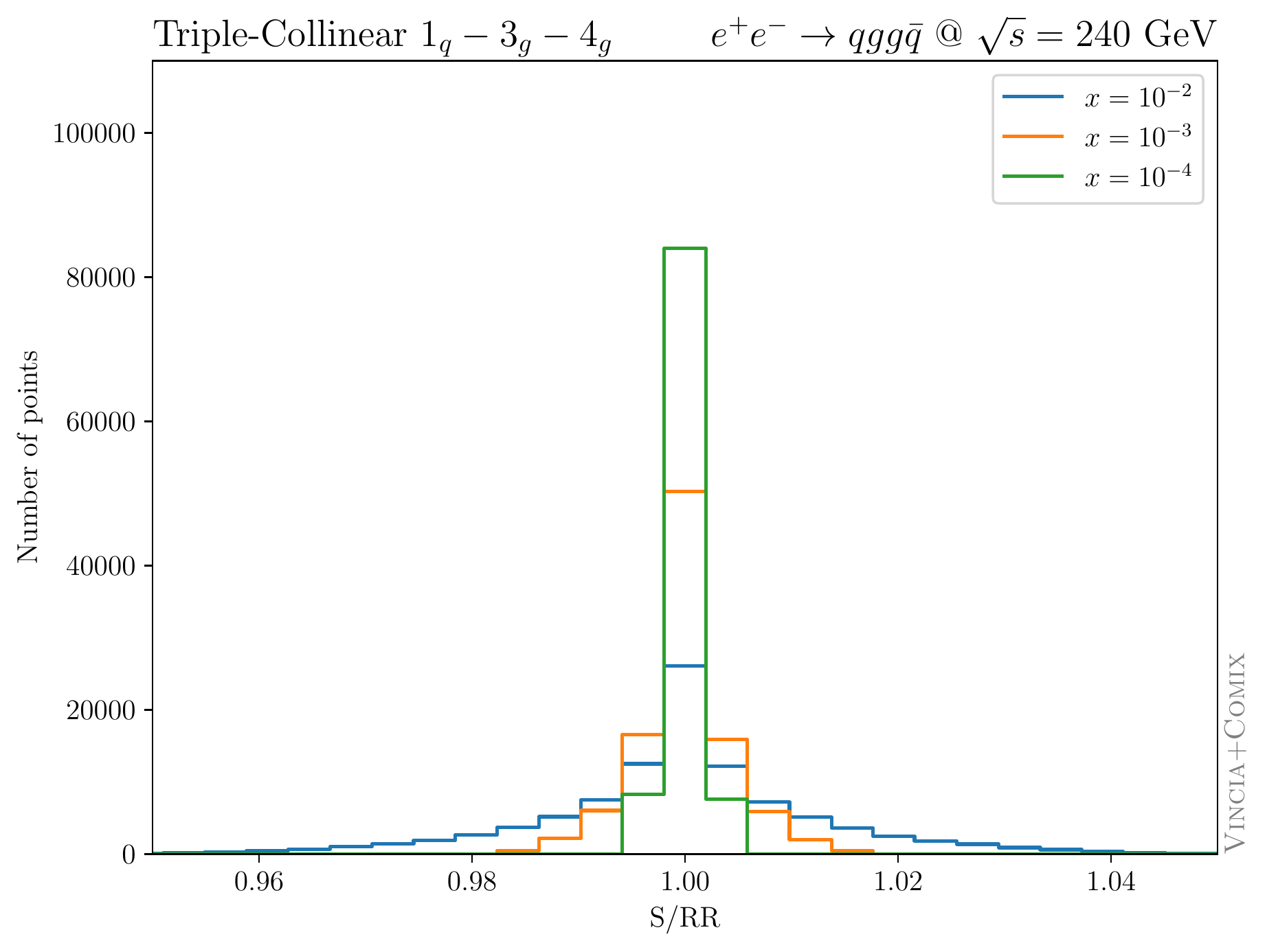}
    \includegraphics[width=0.4\textwidth]{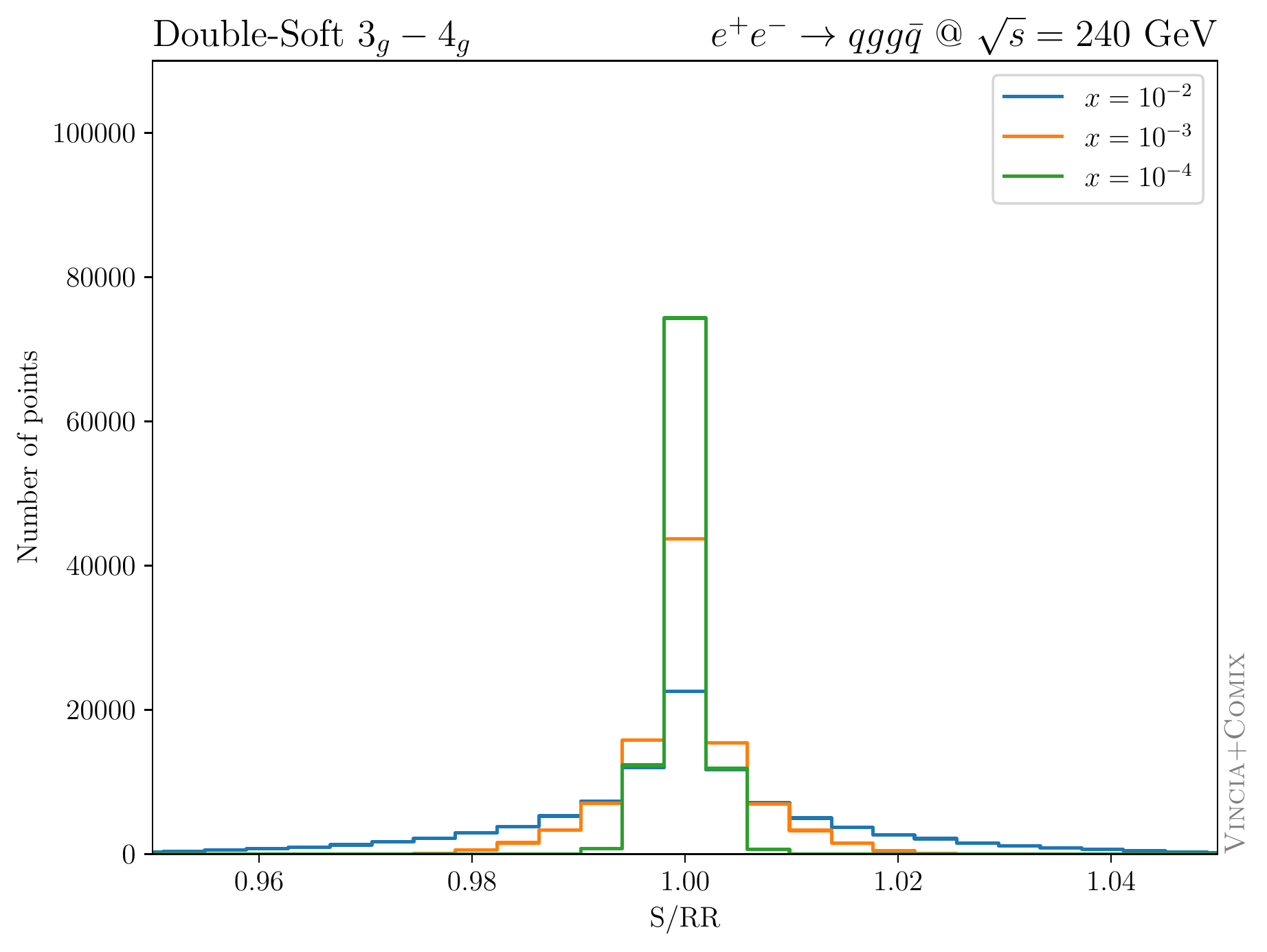}\\
    \includegraphics[width=0.4\textwidth]{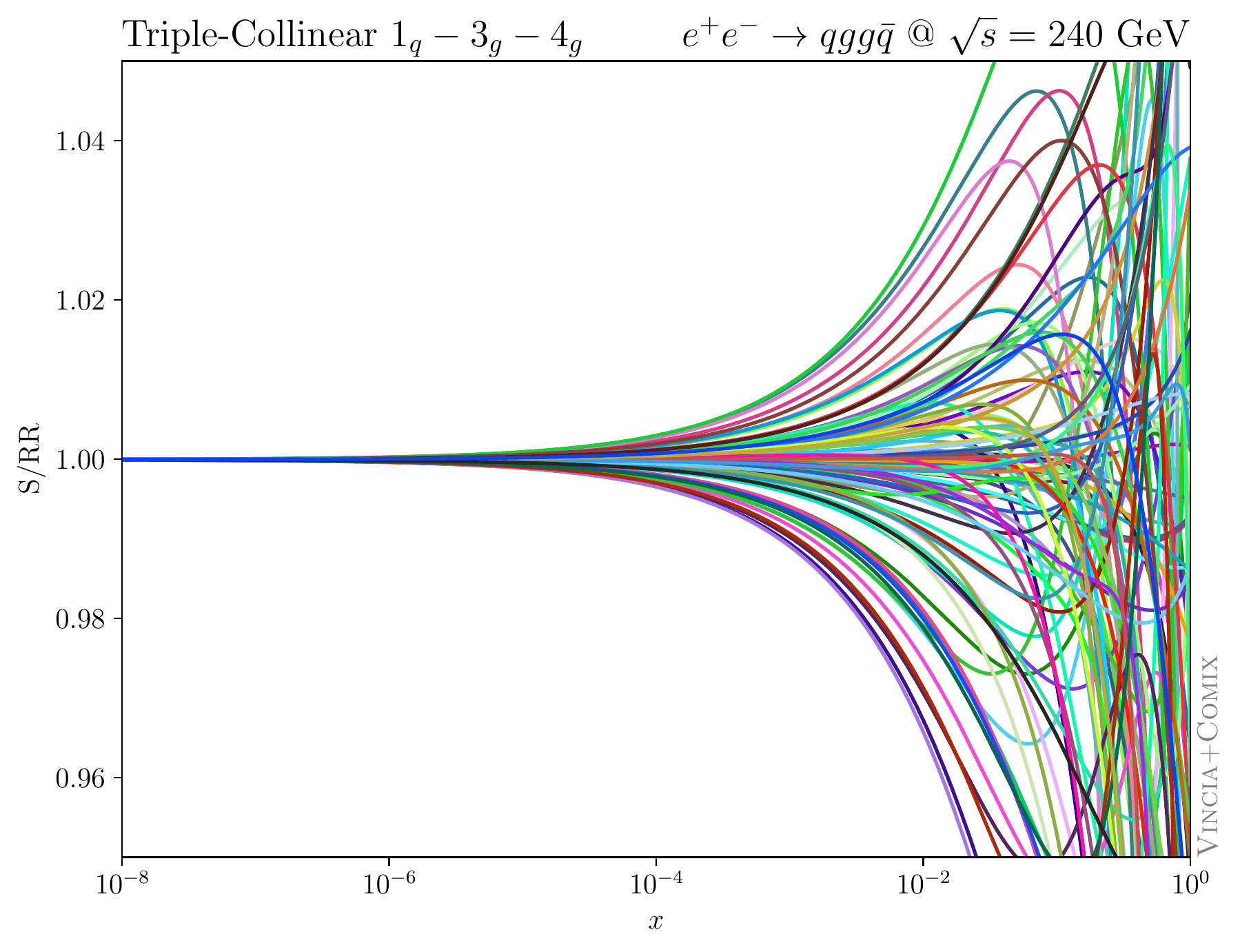}
    \includegraphics[width=0.4\textwidth]{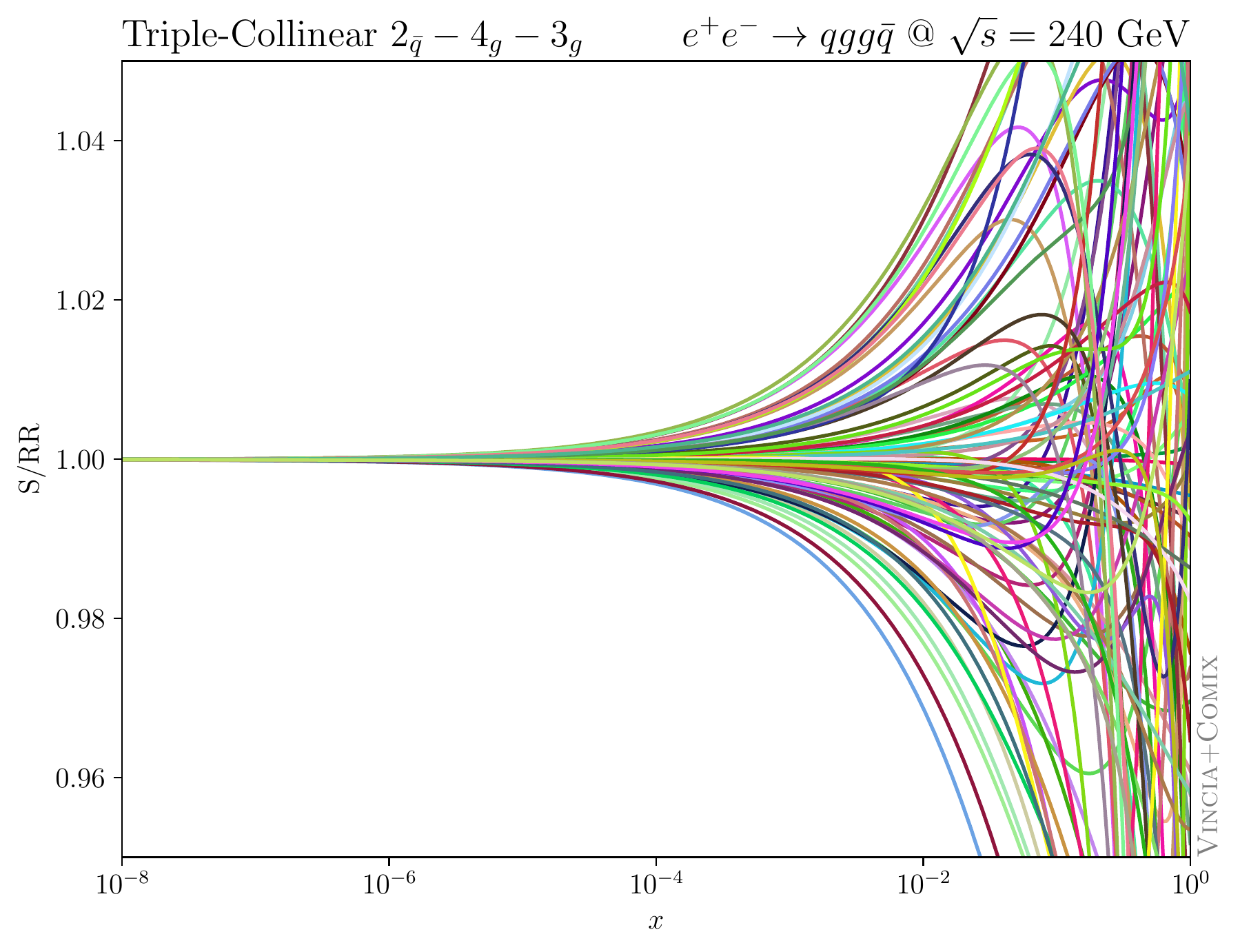}
    \caption{Test of the convergence of the double-real subtraction term $\US(\Phi_{2},\Phi_{+2},O)$ in \cref{eq:SNNLORR} in $e^+e^-\to q g g \bar q$. \textsl{Top row}: progression of weight distributions from $x=10^{-2}$ to $x=10^{-4}$ in the triple-collinear limit ($s_{134}/s_{1234} < x$) and double-soft limit ($s_{134}s_{234}/s_{1234}^2 < x$). \textsl{Bottom row}: trajectories $x\cdot s_{134}$, $x\cdot s_{234}$, $x\to 0$ approaching the two triple collinear limits. Phase space points are not azimuthally averaged.}
    \label{fig:subtractionNNLO}
\end{figure*}

We have implemented the NNLO antenna subtraction terms for processes with two massless final-state jets, cf.~e.g.~\cite{GehrmannDeRidder:2004tv}, in \Vincia in a semi-automated fashion. 
As a validation, we illustrate the convergence of the double-real radiation subtraction term \cref{eq:SNNLORR} in the triple-collinear and double-soft limits for the process $e^+e^-\to q g g \bar q$ in \cref{fig:subtractionNNLO}. Phase space points are sampled according to the kinematic map in \cref{subsec:kinematics} and we do not make use of the azimuthal averaging alluded to above.

It should be noted that a numerical calculation of the Born-local NNLO weight is not necessary for colour-singlet decays, as the inclusive $K$-factors are well known from analytical calculations, cf.~e.g.~\cite{Chetyrkin:1996ela,GehrmannDeRidder:2004tv} for $Z\to q\bar q$ (with massless quarks), \cite{Gorishnii:1990zu,Chetyrkin:1996sr,Baikov:2005rw,DelDuca:2015zqa} for $H \to b\bar b$ (with massless $b$s), and \cite{Chetyrkin:1997iv,GehrmannDeRidder:2005aw} for $H\to gg$ (in the Higgs effective theory).

\section{Conclusions and Outlook} \label{sec:conclusions}
We have presented a technique to match final-state parton showers fully-differentially to next-to-next-to-leading order calculations in processes with two final-state jets. To our knowledge, this is the first method of its kind.

We have outlined a full-fledged numerical implementation in the \Vincia antenna shower in the \Pythia 8.3 event generator. Phenomenological studies employing our strategy will be presented in separate works.

We want to close by noting that, while we here focused on the simplest case of two massless final-state jets, the use of the NNLO antenna subtraction formalism facilitates its adaption to more complicated processes such as $e^+ e^- \to t\bar t$ or $e^+e^-\to 3j$. Considering the latter, spurious singularities in the quark-gluon NNLO antenna subtraction terms need to be removed before exponentiation in the shower.
For future work, an extension of our method to processes with coloured initial states can be envisioned, given the applicability of the NNLO antenna subtraction to hadronic collisions.

\section*{Acknowledgements} 
We thank Aude Gehrmann-de Ridder and Thomas Gehrmann for providing us with FORM files of their antenna functions.
We thank Philip Ilten for the development of a general matrix-element generator interface for \Pythia 8.3, which allowed us to interface \Comix in this work.
CTP is supported by the Monash Graduate Scholarship, the Monash International Postgraduate Research Scholarship, and the J.L.~William~Scholarship.
HTL is supported by the U.S. Department of Energy under Contract No. DE-AC02-06CH11357 and the National Science Foundation under Grant No. NSF-1740142.
This research was supported by Fermi National Accelerator Laboratory (Fermilab), a U.S. Department of Energy, Office of Science, HEP User Facility. Fermilab is managed by Fermi Research Alliance, LLC (FRA), acting under Contract No. DE-AC02-07CH11359.  
This work was further partly funded by the Australian Research Council via Discovery Project DP170100708 — “Emergent Phenomena in Quantum Chromodynamics”.
This work was also supported in part by the European Union’s Horizon 2020 research and innovation programme under the Marie Sk\l{}odowska-Curie grant agreement No 722104 – MCnetITN3.

\bibliographystyle{elsarticle-num}

\bibliography{bibliography.bib}

\end{document}